# Necessity of integral formalism

Yong Tao[†]

School of Economics and Business Administration, Chongqing University, Chongqing 400044, China

**Abstract:** To describe the physical reality, there are two ways of constructing the dynamical equation of field, differential formalism and integral formalism. The importance of this fact is firstly emphasized by Yang in case of gauge field [Phys. Rev. Lett. 33 (1974) 445], where the fact has given rise to a deeper understanding for Aharonov-Bohm phase and magnetic monopole [Phys. Rev. D. 12 (1975) 3845]. In this paper we shall point out that such a fact also holds in general wave function of matter, it may give rise to a deeper understanding for Berry phase. Most importantly, we shall prove a point that, for general wave function of matter, in the adiabatic limit, there is an intrinsic difference between its integral formalism and differential formalism. It is neglect of this difference that leads to an inconsistency of quantum adiabatic theorem pointed out by Marzlin and Sanders [Phys. Rev. Lett. 93 (2004) 160408]. It has been widely accepted that there is no physical difference of using differential operator or integral operator to construct the dynamical equation of field. Nevertheless, our study shows that the Schrödinger differential equation (i.e., differential formalism for wave function) shall lead to vanishing Berry phase and that the Schrödinger integral equation (i.e., integral formalism for wave function), in the adiabatic limit, can satisfactorily give the Berry phase. Therefore, we reach a conclusion: There are two ways of describing physical reality, differential formalism and integral formalism; but the integral formalism is a unique way of complete description.

## 1. Introduction

The quantum adiabatic theorem (QAT) [1-4] is one of the oldest fundamental and most widely used tools in physics. It states that [5] if the Hamiltonian $H(t)$ evolves slowly enough by satisfying the adiabatic condition in time interval $[0,T]$, then the evolving state of the system will remain close to its instantaneous eigenstate up to a multiplicative phase factor in the interval $[0,T]$. In 1984, M. V. Berry [6] found there is a geometrical phase, namely Berry phase, in the adiabatically evolving state vector besides the dynamic phase. Interestingly, Aharonov-Bohm (AB) phase can be regarded as a special case of Berry phase. From that time on, the QAT has been widely accepted by a complete form, that is [7],

[†] Correspondence Email: taoyingyong2007@yahoo.com.cn

$$|\Psi(t)\rangle_A = |\Psi(sT)\rangle_A$$
$$= \exp\left[-iT\int_0^s E_m(s')ds'\right]\exp\left[-\int_0^s \langle m(s')|\frac{\partial}{\partial s'}|m(s')\rangle ds'\right]|m(s)\rangle, \quad (1)$$

where, we have used the scaled dimensionless time variable $s = \frac{t}{T}$.

Recently, however, the QAT has been doubted. Marzlin and Sanders (MS) have pointed out an inconsistency for the QAT [7]. This inconsistency has attracted a lot of attention and has promoted the study of QAT into a new springtide. For this inconsistency, on the one hand, the opinions of critics focus on three points: (i). MS made a mathematical error [8,9,10,11]. (ii). The QAT is physically correct [9,10]. (iii). The conditions for QAT used by MS are insufficient [11,12]. On the other hand, the opinions of supporters focus on two points: (i). Vanishing Berry phase is induced by MS inconsistency [13]. (ii). The widely used quantitative conditions in QAT are insufficient [14,15,16,17,18].

Clearly, there are many different viewpoints as for the inconsistency of QAT raised by MS. Nevertheless, here we need to emphasize: <u>That is because there, in essence, are two different types of inconsistencies of QAT in reference [7]</u>! One is equation (6) in [7] (we call this type the MS inconsistency), another is "counterexample of a two-level system" in [7] (we call this type the MS counterexample). But, it is notable that these two types are often confused with each other by physical circles[1]. In fact, MS inconsistency is almost independent of MS counterexample. Interestingly, the traditional adiabatic condition, $\langle n(t)|\frac{\partial}{\partial t}|m(t)\rangle = 0 \quad (n \neq m)$, neither resolves the MS inconsistency nor removes the MS counterexample. That is the reason why MS inconsistency and MS counterexample are often confused as one. In fact, resolving the MS counterexample is related to references [12,15,16,17,18]; resolving the MS inconsistency is related to references [9,10,13].

Resolving the type of MS counterexample refers to convergence of Schrödinger integral equation in the adiabatic limit, which shows as convergence of transition probabilities[2] between energy levels [17]. Hence, resolving the type of counterexample may guarantee the validity of the adiabatic quantum computation [19]. Fortunately, the type of counterexample has been resolved [12,15,16,17,18]. Recently, Amin [18] gave a splendid summary as for this point. In this paper, we shall not discuss the type of MS counterexample; our purpose is to focus on another type, that is, MS inconsistency. The importance of this inconsistency is still missed by physical circles, except that a handful of references, e.g. [9,10,13], have some substantial contacts.

---

[1] For example, MS inconsistency and MS counterexample are confused as one by references [14] and [15]. In fact, reference [14] is related to the MS inconsistency and reference [15] is related to the MS counterexample. We have clarified this point in a paper [arXiv:1010.1321].

[2] In the rotating representation, the Schrödinger integral equation is written as [15],

$$C_m(sT) = C_m(0) - \sum_{l\neq m}\int_0^s \langle m(s')|\frac{\partial}{\partial s'}|l(s')\rangle \exp\left[iT\int_0^{s'}\omega_{ml}(\sigma)d\sigma\right]C_l(s')ds',$$

where, $\omega_{ml}(\sigma) = E_m(\sigma) - E_l(\sigma)$. Clearly, Convergence of Schrödinger integral equation in the adiabatic limit requires, $\lim_{T\to\infty} C_m(sT) = C_m(0)$. Moreover, the transition probability is determined by $C_m^*(sT)C_m(sT)$.

To resolve the MS inconsistency, here we must pay attention to one important fact: MS reach the MS inconsistency through Schrödinger differential equation [7]. In fact, Wu and Yang [9] have proved that the QAT satisfies Schrödinger integral equation rather than Schrödinger differential equation in the adiabatic limit. They state further that [10] if the QAT satisfies Schrödinger differential equation in the adiabatic limit, then there will be MS inconsistency[3]. Nevertheless, MS, in the reference [7], require that the QAT satisfies the Schrödinger differential equation. That is the reason why there is MS inconsistency in MS's derivation. Another important fact is shown by Pati and Rajagopal [13] that the MS inconsistency shall lead to vanishing Berry phase. The main purpose of this paper is to point out that there is a very important and profound relationship between above two facts; a relationship shows that the Schrödinger differential equation gives rise to the vanishing Berry phase. For this point, we shall provide an explanation: <u>Differential description[4] is not a logical way of describing global effect (or topological effect)</u>. The MS inconsistency just originates from the fact that MS use differential description (i.e., Schrödinger differential equation) to describe a global effect (i.e., Berry phase). To avoid this difficulty, we need to appeal to the integral description.

The organization of our paper is as follows. In section 2, we review the integral formalism for gauge field developed by C. N. Yang and recognize that it is a more proper way of describing AB phase and magnetic monopole than the differential formalism for gauge field. In section 3, we shows that the Schrödinger differential equation would give rise to vanishing Berry phase and that the Schrödinger integral equation, in the adiabatic limit, could satisfactorily give the Berry phase. In section 4, we present the integral formalism for wave function. It can be regarded as a development of integral formalism for gauge field, while it is also an intrinsic requirement of describing Berry phase. In section 5, we analyze the reason why the global effect (e.g., AB phase, magnetic monopole and Berry phase) can be properly described by integral formalism rather than differential formalism. In section 6, we take an example to show how the Schrödinger differential equation, in the adiabatic limit, leads to vanishing Berry phase. In section 7, our conclusion follows. In this paper we set $\hbar = c = 1$.

## 2. AB phase, magnetic monopole, and integral formalism for gauge field.

Before 1960s, it was widely accepted that, for electromagnetic field: (i). Field strength $f_{\mu\nu}$ completely described electromagnetism; (ii). Potential $A_\mu$ were only regarded as a convenient mathematical aid for calculating the fields and hence had no independent significance. However, the study of Aharonov and Bohm [20] shows that $f_{\mu\nu}$ by itself does not, in quantum theory, completely describe all electromagnetic effects on the wave function of electron. They hence suggest that, in quantum mechanics, the fundamental physical entities are the potentials rather than field strength. Later, Yang [21] developed the integral formalism for gauge field and further pointed out that [22] what provides an intrinsic and complete description of electromagnetism is integral formulation $\exp\left(ie\oint A_\mu dx^\mu\right)$.

---

[3] That means, resolving the MS inconsistency, which is different from resolving the type of MS counterexample, but refers to convergence of Schrödinger differential equation in the adiabatic limit.

[4] Here differential description and integral description are determined by differential formalism and integral formalism respectively. The notions of differential formalism and integral formalism shall be presented in footnote (5) of section 2.

From that time on, there have been two ways of describing gauge field [21,22] as follows:

(a). *Differential formalism*[5] *for gauge field*. It is based on the replacement of differential operator $\partial_\mu$ by

$$\partial_\mu - ieA_\mu. \qquad (2)$$

The corresponding Dirac equation reads

$$[\gamma_\mu(i\partial_\mu + eA_\mu) - m]\psi = 0. \qquad (3)$$

(b). *Integral formalism for gauge field*. The wave function of the matter is determined by

$$\Psi(x) = \psi(x)\exp\left[ie\int_P^Q A_\mu dx^\mu\right], \qquad (4)$$

where $\int_P^Q dx^\mu$ denotes any path from $P$ to $Q$ and $\exp\left[ie\int_P^Q A_\mu dx^\mu\right]$ is hence called the nonintegrable (i.e., path-dependent) phase factor.

On the one hand, AB phase can be directly obtained from integral formulation (4) as $P = Q$. Moreover, we can note that AB phase is path-dependent but the phase of wave function satisfying equation (3) is path-independent. Hence, we do realize that the integral formalism (b) is a more natural way of understanding AB phase than the differential formalism (a). In fact, this point has been emphasized by reference [22].

On the other hand, if we not only require that magnetic monopole exists but also require that the Dirac equation (3) holds, then there need to be an additional condition, which reads, $\psi(x) = 0$ along a singular string (Dirac's veto) [23]. Fortunately, this difficulty has been removed by integral formalism (b) [22].

Above discussion shows that the integral formalism (b) is indeed a more natural way of describing global effects (e.g., AB phase and magnetic monopole) than the differential formalism (a). Now that the AB phase is a special case of the Berry phase [6]; a natural question is what provides a more natural way of describing Berry phase, differential formalism or integral formalism? We shall answer this question in the section 4, where we shall surprisingly find that the integral formalism is not only a more natural way of describing Berry phase but is an intrinsic requirement. In other words, the differential formalism can not describe Berry phase! Nevertheless, before entering to this point, we firstly need to explain why Schrödinger differential equation leads to vanishing Berry phase. That is what we shall refer to in next section.

---

[5] Since Maxwell established the electromagnetism theory, people started to believe that the interaction between matters as point contact is passed by fields. In general, field is determined by a dynamical equation. Mathematically, the dynamical equation may be written as a differential equation or an integral equation. In this paper, we call such differential equation the differential formalism and call such integral equation the integral formalism. For example, the matter wave is a field; its differential formalism and integral formalism are Schrödinger differential equation and Schrödinger integral equation respectively. Generally speaking, it has been widely accepted that there is no physical difference of using differential operator or integral operator to construct the dynamical equation of field. Nevertheless, in section 5, we shall find that this viewpoint is wrong.

### 3. MS inconsistency and vanishing Berry phase

Generally speaking, the state vector $|\Psi(sT)\rangle$ of system evolves according to Schrödinger differential equation [24]

$$i\frac{\partial}{\partial s}|\Psi(sT)\rangle = TH(s)|\Psi(sT)\rangle. \quad (5)$$

Clearly, we can note that the Dirac equation (3) is a special case of the Schrödinger equation (5). In general, the QAT reads [4,25] $\lim_{T\to\infty}|\Psi(sT)\rangle = |\Psi(sT)\rangle_A$, where $T \to \infty$ denotes the adiabatic limit [4]. To prove the QAT, an important step is to construct an adiabatic transformation [3,4,24] $\psi(sT)$ which leads to

$$|\Psi(sT)\rangle = \psi(sT)|\phi(sT)\rangle, \quad (6)$$

where, $\psi(sT) = \sum_n \exp\left[-iT\int_0^s E_n(s')ds'\right]|n(s)\rangle\langle n(0)|. \quad (7)$

Here, if we can prove [26]

$$\lim_{T\to\infty}|\phi(sT)\rangle = \exp\left[-\int_0^s \langle m(s')|\frac{\partial}{\partial s'}|m(s')\rangle ds'\right]|m(0)\rangle, \quad (8)$$

the proof of the QAT is complete.

Substitution of equations (6) and (7) into equation (5) gives the dynamical equation of $|\phi(sT)\rangle$ (namely, the Schrödinger equation in the rotating representation),

$$\frac{\partial}{\partial s}|\phi(sT)\rangle = -K_T(s)|\phi(sT)\rangle, \quad (9)$$

or $|\phi(sT)\rangle = |\phi(0)\rangle - \int_0^s K_T(s')|\phi(s'T)\rangle ds'. \quad (10)$

where,

$$K_T(s) = \sum_j \sum_k \exp\left[iT\int_0^s E_j(s') - E_k(s')ds'\right]\langle j(s)|\frac{\partial}{\partial s}|k(s)\rangle|j(0)\rangle\langle k(0)|. \quad (11)$$

Clearly, if we wish to prove the equation (8) (or equivalently, to prove the QAT), we need to solve the equation (9) or the equation (10) in the adiabatic limit $T \to \infty$.

Recently, nevertheless, Wu and Yang [9] have proved that the equation (8), in the adiabatic limit $T \to \infty$, satisfies the integral equation (10) rather than the differential equation (9). Furthermore, they point out that if the equation (8), in the adiabatic limit $T \to \infty$, satisfies the differential equation (9), then there exists a restriction for the Hilbert space,

$$\langle n(s) | \frac{\partial}{\partial s} | m(s) \rangle = 0. \quad (n \neq m) \qquad (12)$$

Indeed, in most cases, there would be no set of $|n(sT)\rangle$ satisfying equation (12) for the Hilbert space. Nevertheless, here we need to emphasize: That is because we, in the past, failed to check the case of adiabatic limit. In fact, the conclusion of reference [9] has shown that if an adiabatic system, in the adiabatic limit, satisfies the Schrödinger differential equation, then the arbitrary eigenstate $|n(s)\rangle$ for the Hilbert space would satisfy the equation (12).

Wu et al, in a later paper [10], continue to prove that the equation (12) shall lead to the MS inconsistency (Detailed proofs sees appendix A), which is

$$\langle m(0) | m(s) \rangle = \exp\left[ \int_0^s \langle m(s') | \frac{\partial}{\partial s'} | m(s') \rangle ds' \right]. \qquad (13)$$

It is easy to check that the equation (12) not only satisfies the traditional adiabatic condition, $\langle n(t) | \frac{\partial}{\partial t} | m(t) \rangle = 0 \ (n \neq m)$, but also is compatible with some new adiabatic conditions given by references [12,15,16,17,18]. Therefore, the conclusion of reference [10] has implied that both the traditional adiabatic condition and these new adiabatic conditions can not still remove the MS inconsistency. However, these new adiabatic conditions may remove the MS counterexample.

Most importantly, Pati and Rajagopal, in reference [13], have proved that the equation (13) shall give rise to vanishing Berry phase. For example, for the cyclic evolution $|m(0)\rangle = |m(1)\rangle$, the equation (13) gives,

$$\exp\left[ \int_0^1 \langle m(s) | \frac{\partial}{\partial s} | m(s) \rangle ds \right] = \exp\left[ \oint \langle m(s) | d | m(s) \rangle \right] = \langle m(0) | m(s) \rangle |_{s=1} = \langle m(0) | m(1) \rangle = 1$$

,

which implies vanishing Berry phase.

Clearly, integrating the conclusions of references [9,10] and [13], we can recognize that the Schrödinger differential equation, in the adiabatic limit, leads to the equation (12) and hence gives rise to vanishing Berry phase. Therefore, we reach a conclusion: The Schrödinger differential equation gives rise to vanishing Berry phase. In fact, we can rigorously prove that if the QAT (1) satisfies the Schrödinger differential equation (5), then the Berry phase vanishes. Proof sees appendix B.

However, the equation (8) satisfies the integral equation (10). That is to say, the Schrödinger integral equation, in the adiabatic limit, can satisfactorily give the Berry phase.

In order to understand this curious situation we are facing, we need to recall that, in the section 2, the AB phase can be satisfactorily described by the integral formalism (b) rather than differential formalism (a) and that the AB phase is a special case of Berry phase. That means, in accordance with the case of gauge field, there are also two different ways of describing wave function. We shall refer to this point in next section.

## 4. Berry phase and integral formalism for wave function

In last section, we have concluded that there are two different ways of describing wave function. That may be the reason why the Berry phase is compatible with the Schrödinger integral equation rather than the Schrödinger differential equation. To show this point, here we attempt to define the clear formulations of these two ways of describing wave function. In analogy with the differential formalism (a) and the integral formalism (b), we define,

(c). *Differential formalism for wave function.* The wave function is determined by the Schrödinger differential equation (9),

$$\frac{\partial}{\partial s}|\phi(sT)\rangle = -K_T(s)|\phi(sT)\rangle.$$

(d). *Integral formalism for wave function.* The wave function is determined by

$$|\Psi(sT)\rangle = \psi(sT)|\phi(sT)\rangle,$$

where $|\phi(sT)\rangle$ is determined by the Schrödinger integral equation (10),

$$|\phi(sT)\rangle = |\phi(0)\rangle - \int_0^s K_T(s')|\phi(s'T)\rangle ds'.$$

Clearly, the integral formalism (d) is, formally, similar to the integral formalism (b)[6]. Further, the integral formalism (b) can be regarded as a special case of the integral formalism (d). For

---

[6] In fact, if we use the Aharonov-Anandan connection [27], the differential formalism for wave function may be also, formally, similar to the differential formalism (a). To understand this point, let $|\Psi(sT)\rangle$ be

$$|\Psi(sT)\rangle = \exp\left[-\int_0^s \langle\omega(s'T)|\frac{\partial}{\partial s'}|\omega(s'T)\rangle ds'\right]|\Omega(sT)\rangle$$

where $\exp\left[-\int_0^s \langle\omega(s'T)|\frac{\partial}{\partial s'}|\omega(s'T)\rangle ds'\right]$, in the cyclic evolution, gives Aharonov-Anandan phase, and

$\langle\omega(sT)|\frac{\partial}{\partial s}|\omega(sT)\rangle$ denotes the Aharonov-Anandan connection. Especially, in the adiabatic limit [27],

$$\lim_{T\to\infty}\exp\left[-\int_0^s \langle\omega(s'T)|\frac{\partial}{\partial s'}|\omega(s'T)\rangle ds'\right] = \exp\left[-\int_0^s \langle m(s')|\frac{\partial}{\partial s'}|m(s')\rangle ds'\right],$$

which, in the cyclic evolution, gives the Berry phase.

Then the differential formalism for wave function is determined by

$$i\left[\frac{\partial}{\partial s}+iTH(s)-\langle\omega(sT)|\frac{\partial}{\partial s}|\omega(sT)\rangle\right]|\Omega(sT)\rangle = 0,$$

which is, formally, similar to differential formalism (a). Especially, in the adiabatic limit, the Aharonov-Anandan connection, $\langle\omega(sT)|\frac{\partial}{\partial s}|\omega(sT)\rangle$, would return to the Berry connection [28],

$\langle m(s)|\frac{\partial}{\partial s}|m(s)\rangle.$

example, the integral formalism (d), in the adiabatic limit, gives,

$$\lim_{T \to \infty} |\Psi(sT)\rangle = \lim_{T \to \infty} \psi(sT) |m(0)\rangle \exp\left[-\int_0^s \langle m(s')| \frac{\partial}{\partial s'} |m(s')\rangle ds'\right], \quad (14)$$

where we have used the equation (8).

Especially, for the cyclic evolution, there may be [6]

$$\exp\left[-\int_0^1 \langle m(s)| \frac{\partial}{\partial s} |m(s)\rangle ds\right] = \exp\left[-\oint \langle m(s)|d|m(s)\rangle\right],$$
$$= \exp\left[ie \oint A_i dx^i\right] \quad (i=1,2,3) \quad (15)$$

which implies that the integral formalism (d) returns to the integral formalism (b).

On the one hand, equations (14) and (15) show that the integral formalism (d), in the adiabatic limit, naturally gives the Berry phase. On the other hand, similar to the AB phase, the Berry phase $\exp\left[-\oint \langle m(s)|d|m(s)\rangle\right]$ is also not integrable; it depends on the path of parameter space [6].

These two facts have implied that the integral formalism (d) is a more natural way of describing Berry phase than the differential formalism (c). Nevertheless, more importantly, the differential formalism (c) can not describe the Berry phase! That is because, according to the discussion of section 3, the Schrödinger differential equation (9) shall give rise to vanishing Berry phase. This point is different from the case of differential formalism (a), which at least, formally, gives the AB phase: That is to say, for any singularity-free paths, the equation (4) satisfies the Dirac equation (3). However, the integral formalism (d), in the adiabatic limit, can satisfactorily give the Berry phase. That is because there, in the adiabatic limit, exists an intrinsic difference between differential formalism (c) and integral formalism (d); a difference leads to that the differential formalism (c) fails to describe Berry phase. We shall present this difference in the next section.

## 5. Shortage of differential formalism

In section 4, we have noted that differential formalism (c) can not describe Berry phase. Nevertheless, it is not really worth surprising in such a fact. That is because, if we carefully check the standard proof of QAT [3,4,24], we shall note that a key step of these proofs is use of Schrödinger integral equation. In other words, we can not reach a complete QAT through Schrödinger differential equation.

Taking an example of the Schrödinger differential equation (9), if we want to reach the equation (8) through the differential equation (9), we need to prove

$$\frac{\partial}{\partial s} \lim_{T \to \infty} |\phi(sT)\rangle = -\lim_{T \to \infty} K_T(s) \lim_{T \to \infty} |\phi(sT)\rangle. \quad (16)$$

To prove the equation (16), we only need to prove

$$\lim_{T \to \infty} \frac{\partial}{\partial s} |\phi(sT)\rangle = \frac{\partial}{\partial s} \lim_{T \to \infty} |\phi(sT)\rangle, \quad (17)$$

where we have used the equation, $\lim_{T \to \infty} K_T(s) |\phi(sT)\rangle = \lim_{T \to \infty} K_T(s) \lim_{T \to \infty} |\phi(sT)\rangle$.

However, the study of reference [9] shows that the equation (17) shall give rise to the equation (12). More importantly, after the discussion of section 3, we have recognized that the equation (12) should lead to the MS inconsistency (13) and that the MS inconsistency (13) should give rise to vanishing Berry phase!

In fact, we are already familiar to an important mathematical fact: Differential does not always commute with limit. That is, commutation relation, $\lim_{T \to \infty} \frac{\partial}{\partial s} = \frac{\partial}{\partial s} \lim_{T \to \infty}$ (e.g., equation (17)), does not always hold. It is neglect of this mathematical fact that leads to MS inconsistency and hence gives rise to vanishing Berry phase.

Similar difficulty also appears in differential description of magnetic monopole. To understand this point, we need to note that the differential is also a limit operation. Gauge invariance of electromagnetic field strength $f_{\mu\nu}$ requires, for a gauge transformation $A_\mu \to A_\mu + \frac{1}{e}\partial_\mu \alpha$, that there holds a commutation relation, $\partial_\mu \partial_\nu \alpha = \partial_\nu \partial_\mu \alpha$. Unfortunately, commutation relation, $\partial_\mu \partial_\nu \alpha = \partial_\nu \partial_\mu \alpha$, does not always hold; otherwise, it shall give rise to a zero magnetic monopole. In fact, only $\oint \partial_\mu \alpha dx^\mu = \oiint (\partial_\mu \partial_\nu - \partial_\nu \partial_\mu) \alpha dx^\mu dx^\nu \neq 0$ gives Dirac's quantization [22].

However, above difficulty does not appear in integral formalism (d). That is because the bounded convergence theorem of Lebesgue [29] would guarantee that integral, in general, commutes with limit (Here we suppose the oscillating factors could guarantee convergence of Schrödinger integral equation in the adiabatic limit). That is to say, commutation relation, $\lim_{T \to \infty} \int_0^s d\sigma = \int_0^s d\sigma \lim_{T \to \infty}$, almost always holds. That is the reason why equation (8) satisfies integral equation (10) rather than differential equation (9) in the adiabatic limit $T \to \infty$. In fact, Berry phase, as a non-trivial topological effect, depends on the existence of the singularity of wave function. Nevertheless, the equation (17) (i.e., $\lim_{T \to \infty} \frac{\partial}{\partial s} = \frac{\partial}{\partial s} \lim_{T \to \infty}$) would rule out any singularity of wave function [30], so the Berry phase vanishes.

Therefore, we indeed do note that, in the adiabatic limit, there exists an intrinsic difference between differential formalism (c) and integral formalism (d) as follow:

In case of differential formalism (c), differential does not commute with adiabatic limit; but in case of integral formalism (d), integral commutes with adiabatic limit.

Since Maxwell wrote down the differential form of Maxwell equation, people always believed that the differential (local) description was a logical way of completely describing physical reality and that there is no physical difference of using differential operator or integral operator to construct the dynamical equation of field. Unfortunately, this viewpoint is wrong. For example, for the topological effect (e.g., Berry phase), there is an important physical difference between the Schrödinger integral equation and the Schrödinger differential equation. Previous discussion has shown that the differential description (e.g., differential formalism (c), or equivalently, the Schrödinger differential equation) can not describe the Berry phase. This fact indicates that the differential description is not a logical way of completely describing physical reality. However, the integral description is a logical way of completely describing physical reality.

## 6. An example of vanishing Berry phase

In section 3, we have pointed out that the Schrödinger differential equation shall lead to vanishing Berry phase. Finally, we use an example to show this fact.

Consider the well known model, a spin-half particle in a rotating magnetic field. The Hamiltonian of this system is

$$H(t) = \mu B \left( \sin\theta \cos\frac{2\pi}{T}t \quad \sin\theta \sin\frac{2\pi}{T}t \quad \cos\theta \right) \cdot (\sigma_x \quad \sigma_y \quad \sigma_z).$$

As is well known, the Berry phase of this system is simply calculated as

$$\gamma(C) = \int_0^1 \langle \uparrow(s) | \frac{\partial}{\partial s} | \uparrow(s) \rangle ds = \oint \langle \uparrow(s) | d | \uparrow(s) \rangle = -\pi(1-\cos\theta) \qquad (18)$$

where, $|\uparrow(s)\rangle$ and $|\downarrow(s)\rangle$ are eigenstates of the Hamiltonian of this system.

Generally speaking, $|\uparrow(s)\rangle$ and $|\downarrow(s)\rangle$ are determined by equations,

$$|\uparrow(s)\rangle = \begin{pmatrix} \cos\frac{\theta}{2} \\ \sin\frac{\theta}{2}\exp(i2\pi s) \end{pmatrix}, \qquad (19)$$

$$|\downarrow(s)\rangle = \begin{pmatrix} -\sin\frac{\theta}{2}\exp(-i2\pi s) \\ \cos\frac{\theta}{2} \end{pmatrix}, \qquad (20)$$

where $\theta$ is an arbitrary real number in interval $[0, \pi]$.

Now we require that this system satisfies the Schrödinger differential equation. Then, according to the discussion of section 3, this system, in the adiabatic limit[7], should satisfy the equation (12), which leads to

$$\langle \uparrow(s) | \frac{\partial}{\partial s} | \downarrow(s) \rangle = 0. \qquad (21)$$

Equation (21) implies that the Schrödinger differential equation, in the adiabatic limit, leads to a new restriction for this system. Now that this system must satisfy the equation (21). Hence, substitution of the equations (19) and (20) into the equation (21) leads to

$$\theta = 0 \text{ or } \pi. \qquad (22)$$

Clearly, equation (22) is the conclusion of equation (21) for this system of a spin-half particle. Now that this system satisfies the equation (21), it also would satisfy the equation (22). Nevertheless, substitution of equation (22) into equation (18) leads to

$$\gamma(C) = -\pi(1-\cos\theta) = 0 \text{ or } -2\pi. \qquad (23)$$

Equation (23) doubtless shows that $\exp[-i\gamma(C)] = 1$, that is, Berry phase vanishes. This fact clearly implies that the equation (21) is redundant. In fact, the complete QAT can be only obtained by using the adiabatic limit $T \to \infty$ and Schrödinger integral equation.

---

[7] Here we need to emphasize: The adiabatic limit can not be neglected, since it is a sufficient condition of leading to Berry phase [6].

## 7. Conclusion

Similar to the fact that there have been two ways of describing gauge field, differential formalism (a) and integral formalism (b); there are also two ways of describing general wave function of matter, differential formalism (c) and integral formalism (d). Most importantly, we find that, for these two ways of describing general wave function of matter, differential formalism (c) is not equivalent to integral formalism (d) in the adiabatic limit. Check such a fact, we note that differential formalism (c) can not describe Berry phase: If we use differential formalism (c) to describe Berry phase, Berry phase would vanish. Fortunately, the integral formalism (d), in the adiabatic limit, can satisfactorily give the Berry phase. Therefore, differential formalism (e.g., Schrödinger differential equation) is though a convenient tool to describe physical reality but is not a complete way. What provides an intrinsic and complete description of physical reality is integral formalism (e.g., Schrödinger integral equation).

## Appendix A

In order to show that the equation (12) leads to the equation (13), here we shall prove two theorems as follows:

**Theorem 1**. In the finite-dimensional Hilbert space, the necessary and sufficient condition, which guarantees that the equation (12) holds, reads

$$\frac{\partial}{\partial s}|m(s)\rangle - \langle m(s)|\frac{\partial}{\partial s}|m(s)\rangle|m(s)\rangle = 0 \qquad (A.1)$$

Proof. Sufficiency $\to$ We construct

$$|\Phi_m(s)\rangle = \frac{\partial}{\partial s}|m(s)\rangle - \langle m(s)|\frac{\partial}{\partial s}|m(s)\rangle|m(s)\rangle. \qquad (A.2)$$

Multiplying $\langle n(s)|$ from the left to the equation (A.2), we get $\langle n(s)|\Phi_m(s)\rangle = 0$ for any integer $n$, where we have used the equation (12).

This implies, in the finite-dimensional Hilbert space, that $|\Phi_m(s)\rangle \equiv 0$.

Necessity $\leftarrow$ Multiplying $\langle n(s)|$ from the left to the equation (A.1), we get

$$\langle n(s)|\frac{\partial}{\partial s}|m(s)\rangle = 0. \quad (n \neq m) \quad \square$$

**Theorem 2**. If the equation (A.1) holds, then there exists the MS inconsistency, that is,

$$\langle m(0)|m(s)\rangle = \exp\left[\int_0^s \langle m(s')|\frac{\partial}{\partial s'}|m(s')\rangle ds'\right].$$

Proof. Multiplying $\langle m(0)|$ from the left to the equation (A.1), we get the differential equation

$$\frac{\partial}{\partial s}\langle m(0)|m(s)\rangle - \langle m(s)|\frac{\partial}{\partial s}|m(s)\rangle\langle m(0)|m(s)\rangle = 0.$$

Integration of this differential equation leads to

$$\langle m(0)|m(s)\rangle = \exp\left[\int_0^s \langle m(s')|\frac{\partial}{\partial s'}|m(s')\rangle ds'\right]. \quad \square$$

Using theorems 1 and 2, we can understand that the equation (12) does lead to the equation (13).

## Appendix B

**Theorem 3**. If the QAT (1) satisfies the Schrödinger differential equation (5) in the adiabatic limit $T \to \infty$, then there exists

$$|m(s)\rangle = \exp\left[\int_0^s \langle m(s')|\frac{\partial}{\partial s'}|m(s')\rangle ds'\right]|m(0)\rangle, \quad (B.1)$$

such that the QAT (1) can be rewritten as

$$|\psi(sT)\rangle = \exp\left[-iT\int_0^s E_n(s')ds'\right]|n(0)\rangle. \quad (B.2)$$

Proof. Substitution of QAT (1) into equation (5) leads to

$$\frac{\partial}{\partial s}\left(\exp\left[-\int_0^s \langle m(s')|\frac{\partial}{\partial s'}|m(s')\rangle ds'\right]|m(s)\rangle\right) = 0.$$

This, clearly, implies that

$$\exp\left[-\int_0^s \langle m(s')|\frac{\partial}{\partial s'}|m(s')\rangle ds'\right]|m(s)\rangle = C_m|m(0)\rangle, \quad (B.3)$$

where $C_m$ is a constant.

By setting $s = 0$ in equation (B.3), we get $C_m = 1$, that is to say, equation (B.1) holds.

Substitution of equation (B.1) into the QAT (1) gives the equation (B.2). $\square$

Most importantly, we note that Berry phase factor vanishes in the equation (B.2).